# Integration of spectral coronagraphy within VIPA-based spectrometers for high extinction Brillouin imaging

*Eitan Edrei,[1] Malte Gather, [2] Giuliano Scarcelli,[1,*]*

[1]*Fischell Department of Bioengineering, University of Maryland, College Park, Maryland 20742 USA*
[2]*SUPA, School of Physics and Astronomy, University of St Andrews, North Haugh, St Andrews KY16 9SS, Scotland, U.K.*
*\*scarc@umd.edu*

**Abstract:** VIPA-based spectrometers have enabled rapid Brillouin spectrum measurements and current designs of multi-stage VIPA spectrometers offer enough spectral extinction to probe transparent tissue, cells and biomaterials. However, in highly scattering media or in the presence of strong back-reflections, such as at interfaces between materials of different refractive indices, VIPA-based Brillouin spectral measurements are limited. While several approaches to address this issue have recently been pursued, important challenges remain. Here we have adapted the design of coronagraphs used for exosolar planet imaging to the spectral domain and integrated it in a double-stage VIPA spectrometer. We demonstrate that this yields an increase in extinction up to 20dB, with nearly no added insertion loss. The power of this improvement is vividly demonstrated by Brillouin imaging close to reflecting interfaces without need of index matching or sample tilting.

## 1. Introduction

For many years, Brillouin light scattering spectroscopy has been used in applied physics and material science to characterize material properties without contact and noninvasively [1-5]. From an experimental standpoint, Brillouin scattering spectroscopy is challenging because it involves the interaction of incident light with acoustic phonons which results in optical frequency shifts on the order of 1–10 GHz (<0.02 nm). Thus, extremely high spectral resolution as well as high spectral extinction are needed to detect weak spontaneous Brillouin signatures next to non-shifted optical signals, which can be more than $10^9$ times stronger [6]. Traditionally, this challenge has been met with multi-pass Fabry-Perot interferometers [7-9]; however, this technique requires long acquisition times of about minutes per spectrum [9]. In recent years, the development of VIPA-based spectrometers [10] dramatically reduced the required acquisition time by parallel spectral detection. This approach enables collecting the entire Brillouin spectrum in one shot and with high throughput efficiency, thus bringing the acquisition times down to ~100 ms, even for low power incident light levels compatible with biological materials [11-15].

The spectral extinction of the various designs of VIPA-based spectrometers has proven sufficient to observe Brillouin signatures of transparent tissues, polymers and biological cells [13, 16-20]. Instead, VIPA-based Brillouin measurements have remained limited in situations where a large amount of non-shifted laser light enters the spectrometer. This occurs when scattering media such as biological tissues are sampled and especially when voxels close to an interface of two materials with a different refractive index are interrogated. This experimental situation is particularly important for *in vivo* Brillouin measurements, where index mismatches cause large back-reflections [21-23]. For example, according to Fresnel equations [24], an interface between two materials with small refractive index difference, e.g. 1.4 vs 1.5, will reflect ~0.12% of the laser light directly into the spectrometer, i.e. a non-shifted component of more than 6 orders of magnitude higher than the Brillouin signal. Without sufficient spectral extinction, such a large non-shifted component bleeds through the rest of the spectral pattern, thus resulting in a strong background at Brillouin-shifted frequencies.

In recent years, a great amount of research has been conducted to increase the spectral extinction of VIPA-based spectrometers. One solution is to increase the spectral selection of the spectrometer, either by adding a VIPA spectrometer stage [25]; or by inserting a narrowband filter within the optical train such as a Fabry-Perot etalon [14, 26] or an absorbing gas chamber [27]. In these cases, though, the improved extinction of 20 dB or more is gained at the expense of significant insertion loss, and/or requires frequency locking to maintain stable rejections. Another approach is to eliminate the non-shifted light component by destructive interference [28]; this can reach up to 35dB added extinction with moderate loss but generally only works for a specific reflection plane and stably maintaining the efficiency of the spectrometer over time can be challenging. Finally, beam shaping of the VIPA output has yielded moderate increase in extinction of up to 10 dB for only 10% insertion loss [13], but the insertion losses become significant if higher extinctions are required [29].

In this paper, we take a different approach to the issue and demonstrate a relatively simple addition to a VIPA spectrometer that can increase extinction by up to 20 dB without reducing the intensity of the Brillouin signal. We found that the remaining obstacle to achieve the ultimate extinction of VIPA spectrometers comes from the diffraction noise of the optical elements within the spectrometer; as a result, we introduced a spectral coronagraphy filter designed to eliminate this noise component. Importantly, our method is compatible with other noise reduction techniques and so could be combined with them in the future to maximize performance.

## 2. Principle

To explain the principle behind our method, we use a double-stage VIPA spectrometer as an example. In each stage of the spectrometer, the VIPA etalon produces a spectrally dispersed pattern in the focal plane of the lens placed just after the etalon (Fig. 1a, planes A and B), i.e. the spectrally-dispersed pattern is the Fourier transform of the electromagnetic field at the output of the VIPA. In a double-stage spectrometer, the two spectral dispersion stages are cascaded orthogonally to each other and the planes of the respective spectrally-dispersed patterns are conjugated (Fig. 1a, planes A and B). The image of the spectrally-dispersed pattern is then projected onto a CCD camera and a pattern like in Figure 1a is formed where the Brillouin signals are surrounded by periodic patterns of non-shifted light components.

In scattering media or close to interfaces, the non-shifted light components include elastically scattered light and laser reflections; thus, they are dominant and the Brillouin-shifted component is difficult to detect. This situation is conceptually similar to what astronomers face when they want to image exosolar planets located near stars, i.e. faint objects located in close proximity of a bright light source. In astronomy, the contrast ratio between the faint object of interest and the bright star is often as high as $10^9$. This has spurred the extensive development of so-called coronagraphs [30-34]. A popular coronagraphy technique was first implemented by Bernard Lyot [34]. In the Lyot coronagraphy technique, the exosolar scene is imaged onto an occulting mask, which is designed to block the light generated by the bright star. However, the diffraction of the light from the bright star due to the finite sized optical elements of the imaging system is often still several orders of magnitude brighter than the faint object of interest. To address this issue, another 4-f imaging system is employed to image the occulting mask plane onto the camera, and a spatial filter is placed in the Fourier plane of the imaging configuration to block high order spatial frequencies generated by the diffraction.

To adapt the coronagraphy solution to Brillouin spectroscopy, we recognized VIPA spectrometers present the opposite experimental situation of exosolar planet imaging: instead of having a bright star in the center of the field of view, several bright peaks are in the corner of the field of view and the faint objects (the Brillouin peaks) sit in the center. The first part of the coronagraphy solution is already implemented in most VIPA spectrometers; two spatial filters, shaped as rectangular masks, are generally employed in the planes of the spectrally dispersed patterns to let Brillouin-shifted signal through while physically blocking all non-shifted light components. These masks effectively block the geometrical path of non-shifted light components; however, they cannot stop high-order spatial frequency components arising from light diffraction within the beam path. This results in strong background noise for the Brillouin signal of interest. To better emphasize the extent of this issue, we consider the typical configuration where a one-inch lens (f=200mm) is used after the VIPAs. The measured spectrum is the convolution of the ideal spectral pattern with an Airy-function given by the diffraction limit of the lens (as illustrated in Fig. 1b). High frequency side lobes typical of the Airy function will thus appear at the boundaries of the blocking masks as illustrated in figure 1c. (Note that Fig. 1b and c employ different intensity scales to better visualize the features in each figure.) Diffraction of the non-shifted component of the light therefore leads to a significant background (intensity up to -35db) in the center of the spectrally-dispersed pattern, regardless of the spectral performances of the VIPA interferometer. Under high back-reflection conditions, this high frequency diffraction pattern can easily overcome the Brillouin signal, as illustrated in figure 1c.

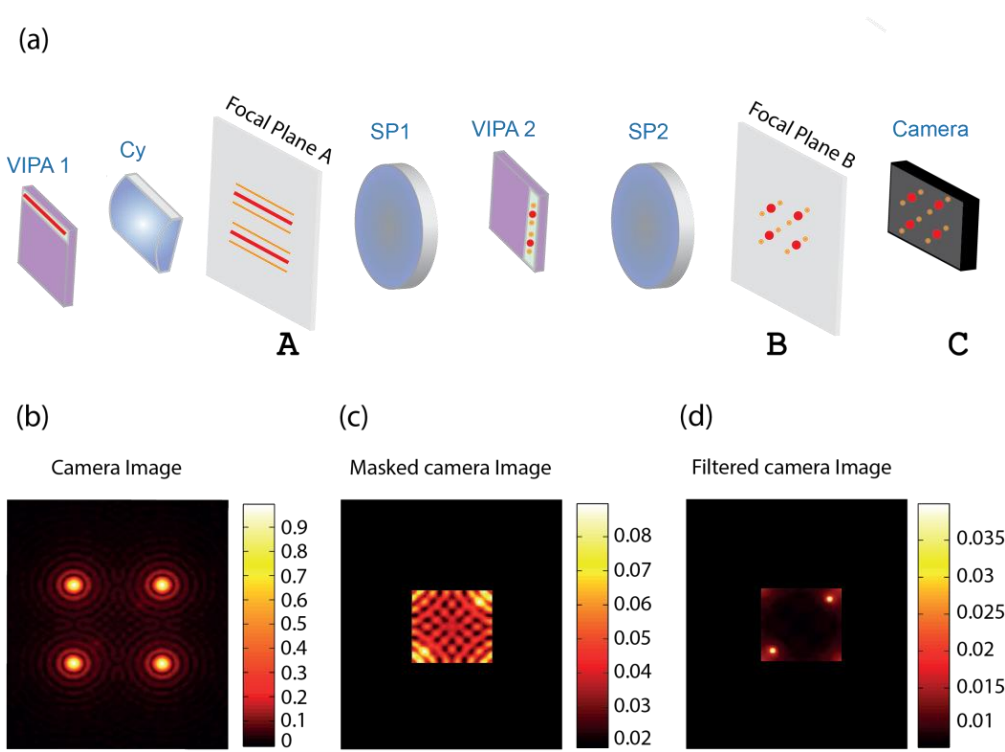

**Figure 1: Illustration of diffraction noise in the double stage VIPA spectrometer:** (a) Schematics of the spectrometer. The first VIPA pattern is observed at the focal plane of a cylindrical lens (plane A) and imaged via a 4-f imaging system and through a second VIPA onto a second plane (plane B). Finally the pattern is recorded by a camera (planes A, B and C are conjugated). (b) An illustration of the Airy patterns and the Brillouin signals as recorded by the camera, under a high back-reflection condition. Two vague Brillouin peaks located within the Airy patterns can be barely distinguished from the background noise. (c) By physically blocking the periphery of the field of view the majority of the light intensity is blocked, yet, the diffraction pattern is still present. (d) In the absence of the diffraction patterns very faint signals can be visualized.

To eliminate the diffraction noise within the spectrometer, similarly to Lyot coronagraphy, another 4-f imaging system can be built to image the plane of the blocking masks onto the camera. A spatial filter is then introduced in the Fourier plane of this imaging system to block the high spatial frequencies generated by diffraction of the strong laser reflections. As illustrated in figure 1d, this is expected to yield a much cleaner spectrum, in which faint signals placed near bright spectral components can be measured.

### 3. Characterization of spectral coronagraphy

To characterize the performance of the coronagraphy filter, we built a Brillouin microscope with a double-stage apodized VIPA spectrometer. The experimental setup is shown in figure 2a. We expanded a single frequency laser beam of wavelength 660 nm (LaserQuantum) and, after a beam-splitter, focused it onto the sample of interest. In epi-detection configuration, we collected the light scattered from the sample and coupled it into a single mode fiber to enter the spectrometer. The spectrometer was composed of two orthogonally oriented VIPAs (LightMachinery, 5mm thickness, 15GHz Free Spectral Range, 95% output reflectivity). We collimated the light leaving the fiber and using a cylindrical lens (Cy1, f=200mm) focused the beam into the VIPA1 etalon tilted in the vertical direction. We used a second cylindrical lens (Cy2, f=200mm) to then focus the output of the VIPA1 etalon and obtain a spectrally-dispersed pattern in the focal plane of Cy2. In this plane, we placed a vertical slit (slit 1) to block the non-shifted laser pattern. Next, we repeated the process similarly using another spectral dispersion stage in the horizontal direction: we focused the pattern transmitted through slit 1 by a spherical lens (SP1, f=200mm) onto the VIPA2 etalon and used a second spherical lens (SP2, f=200mm) to focus the output of VIPA2 and obtain a spectrally-dispersed pattern in the focal plane of SP2. In this plane, we placed a horizontal slit (slit 2) to block the non-shifted laser pattern. As previously mentioned, the second stage of the spectrometer behaves as an imaging system for the vertically-dispersed pattern and the VIPA2 etalon is in the infinity space of the imaging system SP1- SP2. For apodization, after each VIPA etalon we placed a gradient neutral density filter that shaped the VIPA output pattern. The practical implementation of spectral coronagraphy within a

VIPA spectrometer is straightforward: In the final segment of the double staged VIPA spectrometer, we added a 4f system with two identical lenses SP3 and SP4 (f=30mm) and we placed an iris of variable aperture (denoted as "Lyot stop" in Fig. 2) in the Fourier plane which serves as a low-pass spatial filter.

Figure 2b-d illustrate the ability of our method to reduce the diffraction-background noise component. At the interface of a cuvette (n~1.58) and water (n~1.33) the Brillouin signal could not be seen because of large background noise (Fig. 2b), however, by closing the Lyot stop, we were able to clearly observe the Brillouin signal (Fig. 2c). A line average over the width of the signal reveals ~20dB noise reduction compared with the spectrometer with the opened Lyot stop (Fig. 2d).

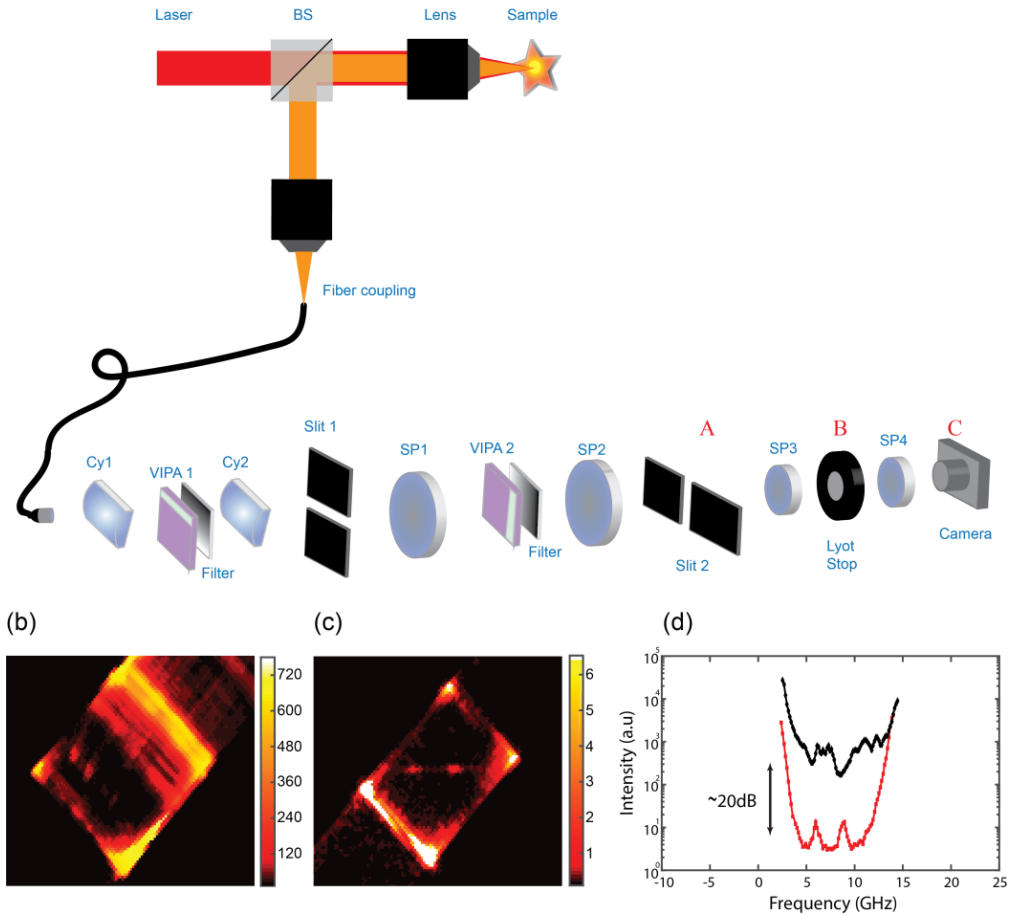

**Figure 2: Setup and quantification of the coronagraphy effect:** (a) An expanded laser beam (red path) is focused into a sample by a 0.7 NA lens (Olympus, LUCPLFLN 60X). Back scattered light (orange path) is collected in an epi-detection configuration and coupled into a single mode fiber. The collimated beam exiting the fiber is focused by a cylindrical lens (Cy1) into the first VIPA; the pattern is gradually filtered and focused onto a vertical slit (slit 1) placed in the focal plane of a second cylindrical lens (Cy2). Next, the pattern is imaged by a 4-f imaging system and through a second VIPA onto a horizontal slit (slit 2). The final plane of the spectrometer (plane A) is imaged onto the camera (plane C) via another 4f imaging system with a spatial filter (Lyot stop) located in the Fourier plane (plane B). (b, c) Brillouin signal from the interface between water and a plastic cuvette, recorded wtih 27mW laser power at the sample and 100ms integration time. The back reflection from the interface was measured to be ~0.7%. (b) With the Lyot stop open the background overcomes the Brillouin signal. (c) By closing the Lyot stop, a clean Brillouin signal can be observed. (d) Average line plot along the line indicated by arrows in b and c, demonstrating a ~20dB noise reduction obtained by the Lyot stop.

The efficiency of noise reduction depends on the diameter size of the Lyot stop relative to the size of the diffracting aperture [35]. To quantify this behavior within our setup, we varied the Lyot stop size and recorded the diffraction-background noise by measuring the light intensity level in the central region of the spectrum between the two Brillouin peaks. Figure 3a shows the experimental results of the Lyot stop diameter reduction (red symbols). For large Lyot stops nearly no reduction of background noise is observed. As the diameter of the Lyot stop decreases, there is a sudden drop of the noise level, followed by a slow monotonic decline. This behavior is expected and can be understood as follows: once the peak of the Airy function is physically blocked by the masks, the remaining portion of the Airy function is a series of diffraction "rings", i.e. intensity peaks and valleys that repeat with a nearly constant period. In the Fourier plane where the Lyot stop is inserted, the diffraction noise therefore has dominant spatial frequencies at a given radius from the center, corresponding to the period of the Airy diffraction rings. As long as the diameter of the Lyot stop lets these dominant spatial frequencies through, the reduction in noise is very small. However, when the diameter of the Lyot stop becomes small enough to block these frequencies, a large drop in diffraction background noise is observed. Once, the dominant noise component has been eliminated, only a moderate further decrease in noise is achieved when closing the Lyot stop further and this corresponds to lower spatial frequency components composing the Airy pattern.

A unique feature of this noise reduction technique is that the measured Brillouin signal is not affected by the blocking aperture placed in the Fourier plane. Figure 3b shows the intensity of the Brillouin signal for different diameter of the Lyot stop. It is evident that within our setup the Lyot stop does not decrease the intensity of the Brillouin signal, even though almost a 20dB rejection of background noise is obtained.

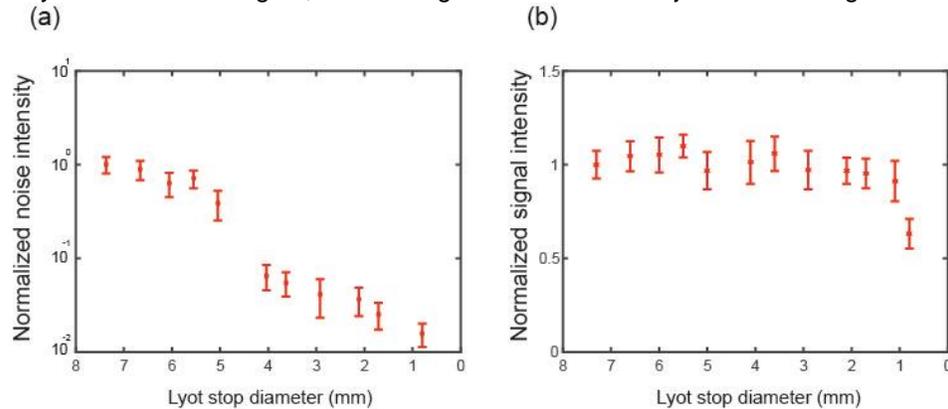

**Figure 3:** (a) Noise levels as a function of Lyot aperture diameter, closing the aperture reduces the noise levels (red symbols). The noise was determined by averaging the value of 30 pixels at the center of the image; the error bars represent the standard deviation over the same region. (b) Brillouin signal intensity as a function of the Lyot stop diameter. For Lyot stop diameters larger than 1mm the signal intensity is not affected. Error bars represent the standard deviation of 25 repeated measurements of the Brillouin peak intensity. All measurements were performed on a water sample as in Fig. 2.

## 4.  Application of coronagraphy to Brillouin imaging

To demonstrate the improvement of our technique in practice, we applied it to Brillouin measurements on a PDMS micro-fluidic channel (width 250µm, height 500µm). We performed Brillouin imaging of the XZ cross-section plane (Fig. 4a) with transverse resolution of 0.5µm and axial resolution of 2µm. The laser intensity at the sample was 25mW and the integration time was set to 0.1sec. We left the channel empty to maximize back-reflection noise due to the refractive index mismatch. The refractive index mismatch between PDMS and air is significant (~0.4); hence, a strong back reflection of 2.8% is expected from each interface plane. The reflected laser light is directed into the spectrometer generating a high level of background noise in the spectrally-dispersed plane recorded by the camera. We used the average intensity value in the position between the Stokes and anti-Stokes Brillouin signals as a measure of the background noise in our system. When the Lyot stop is open, the background collected when recording from a position close to the interfaces between the PDMS and air is more than 200 times higher than the background noise seen in other parts of the sample (Fig. 4b). By closing the Lyot stop, the noise level close at these interfaces was dramatically reduced, by up to 2 orders of magnitude (Fig. 4c).

The Brillouin images of the sample, i.e. maps of the local Brillouin shift, are presented in figure 4d and 4e, showing the measurement acquired with an opened and a closed Lyot stop, respectively. For a fair comparison, both images were processed in the same manner. Pixels located between the two maximal values of every column in figure 4b are shown in black. These pixels were acquired between the two interfaces, and they represent the interior part of the channel that is filled with air and therefore has no Brillouin signal. Importantly, however, when the Lyot stop was open, it was impossible to perform reliable Brillouin imaging of the PDMS material surrounding the air channel, even up to 50 microns away from the air-PDMS interfaces, as shown by the large deviation of Brillouin shift from the expected PDMS values (Fig.

4d). By closing the Lyot stop, we were able to minimize these artifacts, and clear measurements were possible close to the interface between PDMS and air (Fig. 4e).

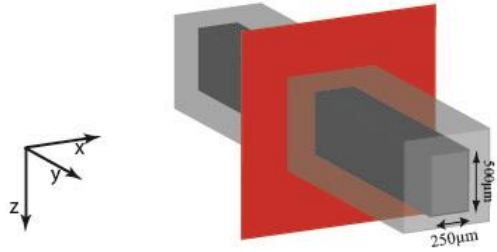

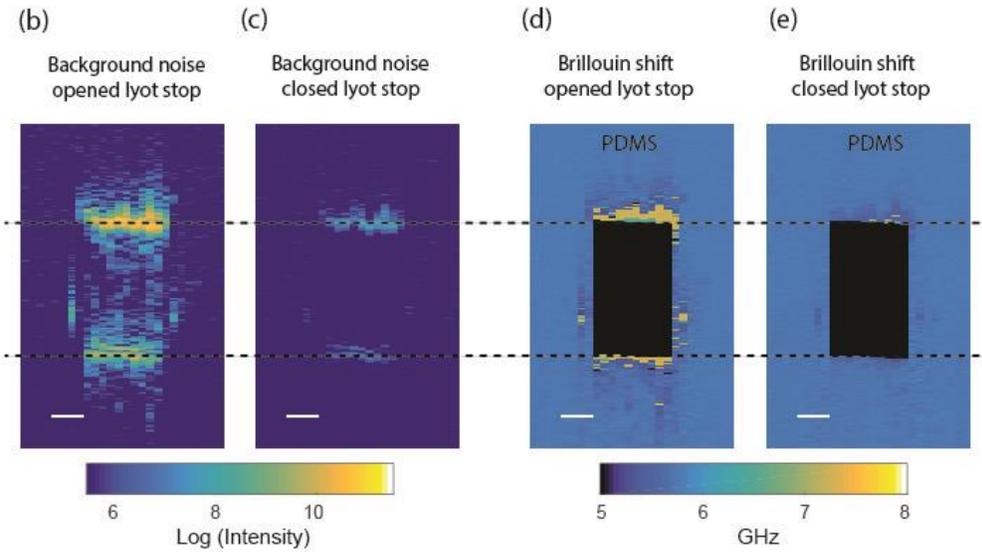

**Figure 4: Brillouin imaging of a micro-fluidic channel:** (a) Illustration of the PDMS micro fluidic channel (width=250µm, height=500µm). Scanning was performed across the XZ plane highlighted in red. Background noise levels defined as the average intensity value between the Brillouin peaks with (b) the Lyot stop widely open and (c) with a closed Lyot stop. The noise from the interfaces was nearly removed for the latter. False color bar indicates camera counts on log scale. Brillouin shifts recorded with (d) an open Lyot stop and (e) a closed Lyot stop. For the former there are clear artifacts up to 50 microns away from the interfaces, whereas the latter show a continuous Brillouin shift up to the interface between air and PDMS.

## 5. Conclusions

In this work, we have introduced the concept of spectral coronagraphy and implemented it within a double stage VIPA spectrometer. The idea addresses a fundamental noise component, which exists in every imaging system, including spectrometers, i.e. the diffraction patterns generated by the finite-sized optical elements. In many scenarios, the diffraction noise element is negligible due to its low intensity, or -- for spectrometers -- it can often be removed by placing a narrow-band filter to block the excitation wavelength. This is the case for example in fluorescence measurements and Raman spectroscopy where the emitted signal and the excitation light are spectrally sufficiently well separated for a standard notch filter to suppress non-shifted light components. However, as demonstrated here, the diffraction noise component can be significant for Brillouin scattering measurements and the frequency shifts involved are too small to be blocked with a conventional notch filter. Intense non-shifted laser light arising from elastic scattering or back-reflections can often occur in the same plane and in close proximity to the low-intensity Brillouin signal of interest, and are thus also immune to confocal rejection or traditional spectral filtering. In this case, a diffraction noise of greater intensity than the measured signal is observed by the spectrometer. We identified this situation to be of great resemblance to what astronomers face when a faint star located close to a bright sun is imaged directly. Hence, we have implemented a similar coronagraphy method within our spectrometer to reduce diffraction noise patterns, and found it to be especially beneficial when the spectrum is acquired close to interfaces between two materials of different refractive index. We quantified the noise reduction in Brillouin imaging settings and found that a rejection of up to ~20dB is achieved with no insertion loss. This will provide important practical advantages for Brillouin imaging. Previous sub-optimal strategies employed to manage the amount of non-shifted light entering the spectrometer (e.g. index matching or sample tilting) thus become largely obsolete. Due to the straightforward implementation of our method, we expect that this technique will find widespread application in VIPA-based spectrometers.

1. R. D. Hartschuh, A. Kisliuk, V. Novikov, A. P. Sokolov, P. R. Heyliger, C. M. Flannery, W. L. Johnson, C. L. Soles, and W. L. Wu, "Acoustic modes and elastic properties of polymeric nanostructures," Applied Physics Letters **87** (2005).
2. J. G. Dil, "Brillouin-scattering in condensed matter" Reports on Progress in Physics **45**, 285-334 (1982).
3. A. A. Serga, T. Schneider, B. Hillebrands, S. O. Demokritov, and M. P. Kostylev, "Phase-sensitive Brillouin light scattering spectroscopy from spin-wave packets," Applied Physics Letters **89** (2006).
4. T. Still, R. Sainidou, M. Retsch, U. Jonas, P. Spahn, G. P. Hellmann, and G. Fytas, "The "Music" of Core-Shell Spheres and Hollow Capsules: Influence of the Architecture on the Mechanical Properties at the Nanoscale," Nano Letters **8**, 3194-3199 (2008).
5. Y. Minami, and K. Sakai, "Ripplon on high viscosity liquid," Review of Scientific Instruments **80** (2009).
6. R. W. Boyd, "Nonlinear Optics, 3rd Edition," Nonlinear Optics, 3rd Edition, 1-613 (2008).
7. J. Sandercock, "Brillouin scattering study of SbSI using a double-passed, stabilised scanning interferometer," Optics Communications **2**, 73-76 (1970).
8. Z. Meng, A. J. Traverso, C. W. Ballmann, M. A. Troyanova-Wood, and V. V. Yakovlev, "Seeing cells in a new light: a renaissance of Brillouin spectroscopy," Advances in Optics and Photonics **8**, 300-327 (2016).
9. F. Palombo, C. P. Winlove, R. S. Edginton, E. Green, N. Stone, S. Caponi, M. Madami, and D. Fioretto, "Biomechanics of fibrous proteins of the extracellular matrix studied by Brillouin scattering," Journal of The Royal Society Interface **11**, 20140739 (2014).
10. G. Scarcelli, and S. H. Yun, "Confocal Brillouin microscopy for three-dimensional mechanical imaging," Nature Photonics **2**, 39-43 (2008).
11. G. Scarcelli, P. Kim, and S. H. Yun, "Cross-axis cascading of spectral dispersion," Optics Letters **33**, 2979-2981 (2008).
12. G. Scarcelli, and S. H. Yun, "In vivo Brillouin optical microscopy of the human eye," Optics Express **20**, 9197-9202 (2012).
13. G. Scarcelli, W. J. Polacheck, H. T. Nia, K. Patel, A. J. Grodzinsky, R. D. Kamm, and S. H. Yun, "Noncontact three-dimensional mapping of intracellular hydromechanical properties by Brillouin microscopy," Nature Methods **12**, 1132-+ (2015).
14. A. Fiore, J. T. Zhang, P. Shao, S. H. Yun, and G. Scarcelli, "High-extinction virtually imaged phased array-based Brillouin spectroscopy of turbid biological media," Applied Physics Letters **108** (2016).
15. G. Scarcelli, R. Pineda, and S. H. Yun, "Brillouin Optical Microscopy for Corneal Biomechanics," Investigative Ophthalmology & Visual Science **53**, 185-190 (2012).
16. K. Elsayad, S. Werner, M. Gallemí, J. Kong, E. R. S. Guajardo, L. Zhang, Y. Jaillais, T. Greb, and Y. Belkhadir, "Mapping the subcellular mechanical properties of live cells in tissues with fluorescence emission–Brillouin imaging," Sci. Signal. **9**, rs5-rs5 (2016).
17. G. Antonacci, and S. Braakman, "Biomechanics of subcellular structures by non-invasive Brillouin microscopy," Scientific Reports **6** (2016).
18. G. Antonacci, R. M. Pedrigi, A. Kondiboyina, V. V. Mehta, R. de Silva, C. Paterson, R. Krams, and P. Török, "Quantification of plaque stiffness by Brillouin microscopy in experimental thin cap fibroatheroma," Journal of the Royal Society Interface **12**, 20150843 (2015).
19. Z. Steelman, Z. Meng, A. J. Traverso, and V. V. Yakovlev, "Brillouin spectroscopy as a new method of screening for increased CSF total protein during bacterial meningitis," Journal of biophotonics **8**, 408-414 (2015).
20. J. Zhang, A. Fiore, S.-H. Yun, H. Kim, and G. Scarcelli, "Line-scanning Brillouin microscopy for rapid non-invasive mechanical imaging," Scientific reports **6** (2016).
21. S. Besner, G. Scarcelli, R. Pineda, and S.-H. Yun, "In Vivo Brillouin Analysis of the Aging Crystalline LensIn Vivo Brillouin Analysis of the Aging Human Lens," Investigative Ophthalmology & Visual Science **57**, 5093-5100 (2016).
22. M. J. Girard, W. J. Dupps, M. Baskaran, G. Scarcelli, S. H. Yun, H. A. Quigley, I. A. Sigal, and N. G. Strouthidis, "Translating ocular biomechanics into clinical practice: current state and future prospects," Current eye research **40**, 1-18 (2015).
23. G. Scarcelli, S. Besner, R. Pineda, P. Kalout, and S. H. Yun, "In Vivo Biomechanical Mapping of Normal and Keratoconus Corneas," Jama Ophthalmology **133**, 480-482 (2015).
24. J. W. Goodman, *Introduction to Fourier optics* (Roberts and Company Publishers, 2005).
25. G. Scarcelli, and S. H. Yun, "Multistage VIPA etalons for high-extinction parallel Brillouin spectroscopy," Optics Express **19**, 10913-10922 (2011).
26. P. Shao, S. Besner, J. Zhang, G. Scarcelli, and S.-H. Yun, "Etalon filters for Brillouin microscopy of highly scattering tissues," Optics Express **24**, 22232-22238 (2016).
27. Z. Meng, A. J. Traverso, and V. V. Yakovlev, "Background clean-up in Brillouin microspectroscopy of scattering medium," Optics express **22**, 5410-5415 (2014).

28. G. Antonacci, G. Lepert, C. Paterson, and P. Torok, "Elastic suppression in Brillouin imaging by destructive interference," Applied Physics Letters **107** (2015).
29. G. Antonacci, S. De Panfilis, G. Di Domenico, E. DelRe, and G. Ruocco, "Breaking the Contrast Limit in Single-Pass Fabry-Pérot Spectrometers," Physical Review Applied **6**, 054020 (2016).
30. P. Baudoz, J. Gay, and Y. Rabbia, "Interfero-coronagraphy: A tool for detection of faint companions," in *Workshop on Brown Dwarfs and Extrasolar Planets*(Puerto La Cruz, Spain, 1997), pp. 254-261.
31. D. Mawet, P. Riaud, O. Absil, and J. Surdej, "Annular groove phase mask coronagraph," Astrophysical Journal **633**, 1191-1200 (2005).
32. F. Roddier, and C. Roddier, "Stellar coronagraph with phase mask," Publications of the Astronomical Society of the Pacific **109**, 815-820 (1997).
33. D. Rouan, P. Riaud, A. Boccaletti, Y. Clenet, and A. Labeyrie, "The four-quadrant phase-mask coronagraph. I. Principle," Publications of the Astronomical Society of the Pacific **112**, 1479-1486 (2000).
34. M. Lyot, "A study of the Solar Corona And Prominences Without Eclipses, 1939," MNRAS **99**, 580.
35. A. Sivaramakrishnan, C. D. Koresko, R. B. Makidon, T. Berkefeld, and M. J. Kuchner, "Ground-based coronagraphy with high-order adaptive optics," Astrophysical Journal **552**, 397-408 (2001).